\documentclass[preprint]{raa}            
\usepackage{graphicx,times}             
\usepackage{natbib}
\usepackage{amssymb,amsmath}
\bibpunct{(}{)}{;}{a}{}{,}

\usepackage[final=true,dvipdfm=true,pagebackref=true]{hyperref}
\hypersetup{colorlinks = true, linkcolor = green, anchorcolor = red, citecolor = blue, filecolor = red, pagecolor = red, urlcolor = red}

\begin{document}

   \title{An explanation about the flat radio spectrum for Mrk 421
}

   \volnopage{Vol.0 (20xx) No.0, 000--000}      
   \setcounter{page}{1}          

   \author{Rui Xue
      \inst{1}
   \and Ze-Rui Wang
      \inst{2}\footnote{Corresponding author}
   }
  \institute{College of Physics and Electronic Information Engineering, Zhejiang Normal University, Jinhua 321004, China\\
  	\and
   School of Astronomy and Space Science, Nanjing University, Nanjing 210093, China; {\it zerui\_wang62@163.com}\\
\vs\no
   {\small Received~~20xx month day; accepted~~20xx~~month day}}

\abstract{It is well known that the flat radio spectrum is a common property in the spectral energy distribution of blazars. Although the one-zone leptonic models are generally successful in explaining the multi-wave band emission, they are problematic in reproducing the radio spectrum. In the study of Mrk 421, one-zone models suggest that in order to avoid overproducing the radio flux, the minimum electron Lorentz factor should be larger than a few hundred at least, even considering the synchrotron self-absorption effect. This result suggests that the model predicted spectral index in the radio band of Mrk 421 should be -1/3. On the basis of this result, by assuming there is a neglected region that will also contribute the radio emission and its electron energy index is naturally originate from the simplest first-order Fermi acceleration mechanism, we can get a superimposed flat radio spectrum. In this paper, a two-zone model is proposed to reproduce the quiescent state spectral energy distribution of Mrk 421. In addition to taking into account the emission from a conventional radiation zone, we further consider the emission from the acceleration zone in which particles are accelerated at a shock front. With the present model, our fitting result suggests that the low frequency flat radio spectrum of Mrk 421 might be explained as a superposition of the synchrotron emission from acceleration zone and radiation zone. 
}

   \authorrunning{Xue, Rui \& Wang, Ze-Rui}            
   \titlerunning{An explanation about the flat radio spectrum}  

   \maketitle

%
%
\section{Introduction}

Blazars, including flat-spectrum radio quasars (FSRQs) and BL Lacertae objects (BL Lacs), are radio-loud active galactic nuclei (AGN) observed with their relativistic jet axis along the line of sight \citep{1995PASP..107..803U, 1997ARA&A..35..445U}. The FSRQs and BL Lacs are defined according to the presence or absence of strong broad-emission lines \citep{1995PASP..107..803U, 1997A&A...325..109S}. The broadband spectral energy distribution (SED) of blazars are quite well studied over all accessible energy bands \citep{2016ARA&A..54..725M, 2020ApJS..248...27T}. In the log${\nu}$-log${\nu}F_{\nu}$ diagram, the typical SED of a blazar displays a double hump structure. In the leptonic model, low energy hump is generally accepted to derive from the synchrotron emission of relativistic electrons in the jet, and high energy hump is derived from inverse Compton (IC) scattering. However, the origin of the scattered photons is not yet clearly identified. If soft photons are derived from synchrotron emission, the IC scattering is called synchrotron self-Compton (SSC) process \citep{1967MNRAS.137..429R, 1974ApJ...188..353J, 1985ApJ...298..114M, 1992ApJ...397L...5M, 1994ApJ...421..153S, 1996ApJ...461..657B}; If soft photons are derived from exterior of jets, the IC scattering is called external Compton process (EC). In the latter case, soft photons can be produced directly by the accretion disc \citep{1992A&A...256L..27D, 1993ApJ...416..458D} or indirectly, for instance, those reprocessed by the broad line region \citep{1994ApJ...421..153S, 1997ApJS..109..103D}, or by the dust torus \citep{2000ApJ...545..107B, 2002A&A...386..415A}. In hadronic models, the high-energy bump originates from proton-synchrotron emission or emission from secondary particles \citep{2000NewA....5..377A, 2019ApJ...871...81X}, and the generated high-energy neutrino emission might be observed \citep[e.g.,][]{2019ApJ...886...23X}.

Observationally, there is a break (${\nu}_{\rm break}$$\sim$$10^{11}$Hz) in the standard radio spectrum of blazars. In the high frequency region ($\nu > {\nu}_{\rm break}$), the radio spectrum steepens, signaling that the jet emission is optically thin. In the low frequency region ($\nu < {\nu}_{\rm break}$), most blazars show a flat spectrum, which is likely to be self-absorbed (thus optically thick) \citep{1979ApJ...232...34B, 1985A&A...146..204G, 2009arXiv0909.2576M}. On the other hand, there are some blazars that do not have significant break in the radio spectrum \cite[e.g., see the SEDs of J0006-0623, J0102+5824, J0433+0521 in][]{2011A&A...536A..15P}. Statistically, the distributions of low frequency spectral indices are similar for FSRQs and BL Lacs with ${\alpha}$$\sim$-0.1, which mean that the flat radio spectrum is a common property for both types of blazars \citep{2011A&A...536A..15P, 2012A&A...541A.160G}.

In order to explain this flat radio spectrum, many studies make efforts to reproduce it with inhomogeneous model \citep{1979ApJ...232...34B, 1980ApJ...235..386M, 1981ApJ...243..700K, 2006MNRAS.367.1083K, 2009ApJ...699.1919P, 2012MNRAS.423..756P, 2013MNRAS.429.1189P, 2013MNRAS.431.1840P, 2013MNRAS.436..304P, 2015MNRAS.453.4070P}. A jet model with a uniform conical structure was firstly proposed by \cite{1979ApJ...232...34B}. Assuming that the magnetic luminosity is conserved in each segment, the flat radio spectrum is reproduced by a superposition of synchrotron self-absorbed radiation from each segment along the jet. However, the adiabatic losses are treated as an optional presence in the standard Blandford-K{\"o}nigl jet model. When considering adiabatic loss, the re-acceleration of electrons must exist so that the adiabatic energy loss rate is balanced by the electron re-acceleration rate \citep{2006MNRAS.367.1083K, 2013MNRAS.429.1189P, 2015MNRAS.453.4070P, 2019MNRAS.485.1210Z}, otherwise the flat radio spectrum remains unexplained.

From the perspective of the one-zone homogeneous model which is widely used in blazar research, although the SEDs of Blazars can be reproduced successfully \citep{2011MNRAS.414.2674G, 2011ApJ...742...27L, 2012ApJ...752..157Z, 2014MNRAS.439.2933Y, 2014ApJ...788..104Z, 2018ApJS..235...39C, 2020ApJS..248...27T}, the flat radio spectrum cannot be fitted in most cases. From the SEDs that fitted by the one-zone models, one can find that the observed radio flux is higher than the model predicted flux which considered self-absorption. Many studies suggest that the low frequency flat radio emission must come from a large-scale jet which is a larger and less compact region, therefore the one-zone homogeneous leptonic model that is aimed to explain the compact jet radiation cannot explain the low frequency radio emission \citep{2009MNRAS.399.2041G, 2010MNRAS.402..497G}. However such explanations may be arbitrary, the extended radiation from large-scale jets is different from the beamed core emission, and usually exhibits a steep spectrum at low frequencies \cite[e.g., see Fig. 1 in][]{2011ApJ...740...98M}. In radio observation, the observed steep spectrum, rather than the flat spectrum, is also used as the basis for judging whether the emission comes from a large-scale area \citep[e.g.,][]{2015AJ....149...46Z}. Therefore it cannot be simply considered that the unexplained flat radio spectrum of one-zone model comes from the extended large-scale jet.

In this paper, without considering the emission from extended region, we investigate the possibility that the emission at flat spectrum could also come from the acceleration region, which is not discussed in previous studies. In this case, even if the synchrotron emission of the radiation region is dominant in low energy component, the acceleration region still has the potential to contribute the emission at radio band. Therefore, we propose a two-zone model in which particles are accelerated at a shock front and cooled via synchrotron and IC radiation in a homogeneous magnetic field behind it. In the following, we will discuss whether the radiation from the acceleration zone could contribute the low frequency flat radio spectrum of a well-known blazar, Mrk 421. This paper is structured as follows. In Sect.~\ref{model} we present the model description; in Sect.~\ref{app} we model the SED of Mrk 421; in Sect.~\ref{dis} we provide some discussions and conclusions. 

\section{Model description}\label{model}

\begin{figure}
\includegraphics[width=90mm]{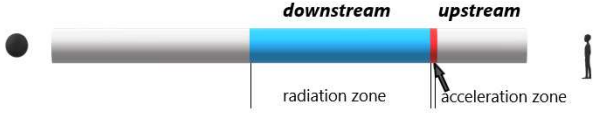}
\caption{A sketch, not to scale, for the two-zone model.} \label{sket}
\end{figure}

A two-zone scenario is proposed to explain the broadband emission in this paper. In the model, we basically follow \cite{1998A&A...333..452K} and reconsidered the emission from the acceleration zone. We assume that the observed SED is a superposition of two zones that are radiating contemporaneously and boosted with the same Doppler factor ($\delta$). We treat two zones: one around the shock front where the electrons are continuously accelerated from an initial value of the electron Lorentz factor ${\gamma}_{\rm 0}$ up to ${\gamma}_{\rm max}$ and then escape into a downstream region where electrons emit most of the radiation with an energy-independent rate $t_{\rm esc}^{-1}$ . We assume a cylindrical geometry with the same constant cross-section for the two zones, and the length of the acceleration zone $L_{\rm acc}$ is relatively thinner than the length of the radiation zone $L_{\rm rad}$ (see Fig.~\ref{sket}). A brief description of the two-zone model is provided as follows \cite[for more details, see][]{1998A&A...333..452K}.

Summing up all energy gain and loss terms, electrons injection and escape terms, we get the kinetic equation that governs the evolution of electrons in the acceleration zone
\begin{equation}\label{acc}
\frac{\partial N_{\rm acc}(\gamma , t)}{\partial t} + \frac{\partial}{\partial \gamma}[(\frac{\gamma}{t_{\rm acc}} - \beta_{\rm s} \gamma^2)N_{\rm acc}(\gamma , t)] + \frac{N_{\rm acc}(\gamma , t)}{t_{\rm esc}} = Q \delta(\gamma - {\gamma}_{0}),
\end{equation}
where $N_{\rm acc}(\gamma , t)$ is the electron number density per $\gamma$ in the acceleration zone, $\gamma$ is the electron Lorentz factor, ${t}_{\rm acc}$ is the characteristic time for the shock acceleration gains, $\beta_{\rm s}\gamma^2 = \frac{4\sigma_{\rm T}B^2\gamma^2}{3m_{\rm e}c8\pi}$ describes the synchrotron losses, $Q \delta(\gamma - {\gamma}_{\rm 0})$ describes the injection of monochromatic electrons of energy $\gamma_{\rm 0}$ into the acceleration process, $\delta(\gamma - {\gamma}_{\rm 0})$ is the $\delta$-function.

For $t > 0$ and $\gamma_{\rm 0} <  \gamma < \gamma_{\rm 1}(t)$ , the solution of Eq.~\ref{acc} is given by
\begin{equation}
N_{\rm acc}(\gamma, t) = a\frac{1}{\gamma ^2}(\frac{1}{\gamma} - \frac{1}{\gamma _{\rm max}})^{(t_{\rm acc}-t_{\rm esc})/t_{\rm esc}},
\end{equation}
where
\begin{equation}
\gamma_1(t) = (\frac{1}{\gamma_{\rm max}}+[\frac{1}{\gamma_0} - \frac{1}{\gamma_{\rm max}}e^{-t/t_{\rm acc}}])^{-1}
\end{equation}
with $\gamma_{\rm max} = (\beta_{\rm s}t_{\rm acc})^{-1}$, and
\begin{equation}
a = Q_0t_{\rm acc}\gamma_0^{t_{\rm acc}/t_{\rm esc}}(1-\frac{\gamma_0}{\gamma_{\rm max}})^{-t_{\rm acc}/t_{\rm esc}},
\end{equation}
where ${Q}_{0}$ is a constant injection rate after switch-on at time t=0. $t_{\rm acc}$ and $t_{\rm esc}$ are assumed to be independent of energy here.

After being accelerated, electrons leave the acceleration zone at a rate $N_{\rm acc}$($\gamma, t$)/$t_{\rm esc}$ and enter the radiation zone where they loss most of their energy via synchrotron and IC process. The evolution of electrons in the radiation zone is governed by the following kinetic equation
\begin{equation}\label{rad}
\frac{\partial N_{\rm rad}(\gamma , t)}{\partial t} - \frac{\partial}{\partial \gamma}((\beta_{\rm s} \gamma^2 + P_{\rm IC}) N_{\rm rad}(\gamma , t)) = \frac{N_{\rm acc}(\gamma , t)}{t_{\rm esc}} \delta(x - x(t)),
\end{equation}
where $N_{\rm rad}(\gamma , t)$ is the electron number density per $\gamma$ in the radiation zone, $P_{\rm IC} = m_{\rm e}^3c^7h\int_{0}^{\xi_{\rm max}}d\xi \xi\int_0^{\infty}d\xi_1N_{\rm ph}(\xi_1)\frac{N(\gamma, \xi_1)}{dtd\xi}$ describes the IC losses including the Klein-Nishina effect \citep{2002cra..book.....S}, where $N_{\rm ph}$ is the differential photon number density, $\xi$ and $\xi_1$ are the non-dimentional energies for the scattered photons and the target photons, ${x}(t)$ is the position of shock front at time $t$, $\delta(x - x(t))$ is the $\delta$-function. Note that the acceleration term of Eq.~\ref{acc} is not considered here, since electrons only get accelerated in the acceleration zone in this model. 

The kinetic equations for acceleration zone and radiation zone are solved for the time, space and energy dependences of the particle distribution function. Therefore, this two-zone model is homogeneous in the sense that the magnetic field does not vary, but contains an inhomogeneous electron energy distribution. On the other hand, by assuming the magnetic field intensity B for the two zones is the same, we can calculate their synchrotron emission coefficients through
\begin{equation}
j_{\rm syn}(\nu) = \frac{1}{4\pi} \int N_{\rm acc, rad}(\gamma)P(\nu, \gamma) d\gamma,
\end{equation}
where P($\nu$, $\gamma$) is the mean emission coefficient for a single electron integrated over the isotropic distribution of pitch angles. And the synchrotron absorption coefficient for the acceleration zone and the radiation zone are calculated with
\begin{equation}
\alpha_{\rm syn}(\nu) = -\frac{1}{8\pi \nu^2 m} \int d\gamma P(\nu, \gamma)\gamma^2 \frac{\partial}{\partial \gamma}[\frac{N_{\rm acc, rad}(\gamma)}{\gamma ^2}].
\end{equation}
Then we can calculate the synchrotron intensity using the radiative transfer equation
\begin{equation}
I_{\rm syn}(\nu) = \frac{j_{\rm syn}(\nu)}{\alpha_{\rm syn}(\nu)}(1-e^{-\alpha_{\rm syn}(\nu) L_{\rm acc, rad}}),
\end{equation}
where $\alpha_{\rm syn}(\nu) L_{\rm acc, rad}$ is the optical depth.

The SSC emission coefficient is given by
\begin{equation}
j_{\rm SSC} = \frac{h\epsilon}{4\pi} \int d\epsilon_0 n(\epsilon_0) \int \gamma N_{\rm acc, rad}(\gamma)C(\epsilon, \gamma, \epsilon_0),
\end{equation}
where $\epsilon$ is the scatted photon energy, ${\epsilon}_{0}$ is the soft photon energy. C($\epsilon$, $\gamma$, ${\epsilon}_{0}$) is the Compton kernel given by \cite{1968PhRv..167.1159J}
\begin{equation}
C(\epsilon, \gamma, \epsilon_0) = \frac{2\pi r_{\rm e} ^2 c}{\gamma^2 \epsilon_0}[2\kappa ln(\kappa) + (1 + 2\kappa)(1 - \kappa) + \frac{(4\epsilon_0 \gamma \kappa)^2}{2(1+4\epsilon_0 \gamma \kappa)}(1 - \kappa)],
\end{equation}
where
\begin{equation}
\kappa = \frac{\epsilon}{4\epsilon_0 \gamma (\gamma - \epsilon)},
\end{equation}
and n(${\epsilon}_{0}$) is the number density of the synchrotron photons per energy interval.

Because the medium for the SSC radiation field is transparent, we can calculate the SSC intensity
\begin{equation}
I_{\rm SSC}(\nu) = j_{\rm SSC}(\nu)L_{\rm acc, rad}.
\end{equation}
Then we can get the total observed flux density as:
\begin{equation}
F_{\rm obs}(\nu_{\rm obs}) = \frac{\pi R^2 \delta^{3}(1+z)}{D_{\rm L} ^2} (I_{\rm syn}(\nu) + I_{\rm SSC}(\nu)),
\end{equation}
where ${D}_{\rm L}$ is the luminosity distance, $z$ is the redshift and ${\nu}_{\rm obs}$ = $\nu \delta/(1+z)$, $R$ is the radius of the cross-section. In addition, the very high energy (VHE) $\gamma$-ray photons will be absorbed by the extragalactic background light (EBL). It makes the observed spectrum in the VHE band steeper than the intrinsic one. According the EBL model that presented by \cite{2011MNRAS.410.2556D}, we calculate the absorption in the GeV-TeV band.

\section{Application}\label{app}
\subsection{Modeling the SED of Mrk 421}\label{Modeling}
The high synchrotron peak BL Lac Mrk 421, at a redshift of $z$ = 0.03, is one of the brightest TeV blazars that has been well studied by many researchers \citep[e.g.][]{2011ApJ...736..131A}. In the study of its SED, the conventional one-zone leptonic model suggest that in order to avoid predicting a larger radio flux than the observed data, the minimum Lorentz factor $\gamma_0$ should be much larger than the typical value which is in the range between 1 and 100 \citep{2008MNRAS.385..283C, 2014ApJ...788..104Z}. In the one-zone SSC model, \cite{2011ApJ...736..131A} set $\gamma_0 = 400, 800$, \cite{2014A&A...567A.135A} set $6000 < \gamma_0 < 17000$ and \cite{2014MNRAS.439.2933Y} set $\gamma_0 = 500$. This implies that if $\gamma_0$ is large enough, the spectral index $\alpha_1$ ($F_\nu \propto \nu^{-\alpha_1}$) in the radio band should be -1/3 which contributed by the monochrome electron $\gamma_0$. Moreover, if the acceleration zone also can contribute to the radio emission and the electrons is accelerated through the simplest first-order Fermi acceleration mechanism, it will give a canonical value for $\alpha_2$ about 0.75 \citep{2015SSRv..191..519S}. Then we can obtain a superimposed flat radio spectrum with $\alpha \approx 0.2$ ($\alpha \approx (\alpha_1 + \alpha_2)/2$). Motivated by the above issues, in this paper, we employ a two-zone model that presented in Sect.~\ref{model} to reproduce the broadband SED of Mrk 421 and investigate whether the acceleration zone could contribute flux in the radio band and stack with the emission of radiation zone to get the low frequency flat radio spectrum. The averaged SED of Mrk 421 resulting from quasi-simultaneous observations integrated over a period of 4.5 months is provided by \cite{2011ApJ...736..131A}, which is the most complete SED ever collected for Mrk 421. This averaged SED is a good proxy for the quiescent state because of its low activity and relatively low variability \citep{2010ASPC..427..277P}.

\begin{figure}
\includegraphics[width=84mm]{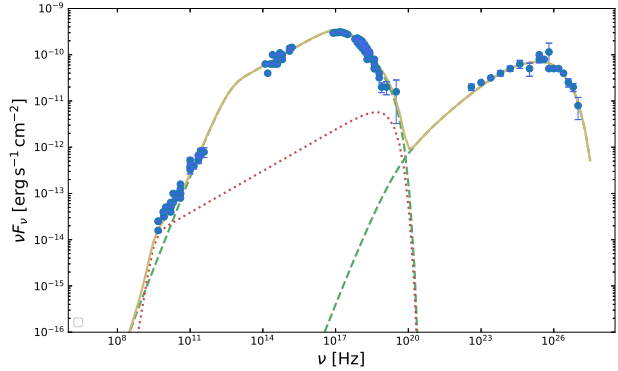}
\caption{The two-zone SSC model for SED modeling of Mrk 421. The red dotted line represents synchrotron emission in acceleration zone. The green dashed line represents synchrotron and SSC emission in radiation zone. The yellow solid line represents the total spectrum by summation both of the acceleration zone and radiation zone emission.} \label{sed}
\end{figure}

There are 11 parameters in the two-zone model, which are $Q_0$, $R$, $L_{\rm acc}$, $L_{\rm rad}$, $B$, $\delta$, $t_{\rm acc}/t_{\rm esc}$, $t_{\rm obs}$, $\gamma_{\rm 0, acc}$, $\gamma_{\rm 0, rad}$ and $\gamma_{\rm max}$, respectively. As introduced in Sect.~\ref{model}, $L_{\rm acc}$ and $L_{\rm rad}$ would affect the synchrotron and SSC intensities, and $L_{\rm acc}$ should be relatively thinner than $L_{\rm rad}$. In this work, the acceleration mechanism is not specified, but we assume the Fermi-type acceleration mechanism works. In order to accelerate the particles effectively, the mean free path between clouds of plasma should be smaller than $L_{\rm acc}$. Here, we further assume that $L_{\rm acc}$ is ten times the mean free path between clouds of plasma. The mean free path between clouds of plasma can be estimated as $L = \frac{4v^2t_{\rm acc}}{3c}$ \citep{1995ApJ...446..699P}, where $t_{\rm acc} = 1/(\gamma_{\rm max}\beta_{\rm s})$ and $v = 0.1c$ is the velocity of clouds of plasma relative to the particles. The model is not sensitive to the parameter $\gamma_{\rm max}$, so we set $\gamma_{\rm max} = 2\times10^6$ in this work. Then we can get $L = 2.16\times 10^{15}$cm and $L_{\rm acc} = 2.16\times 10^{16}$cm. Following previous studies, we set $\gamma_{\rm 0, acc} \approx 48$ in the modeling \citep{2014ApJ...788..104Z}. If assuming that the upstream electrons moving in the opposite direction with an upstream Lorentz factor $\gamma_{\rm 0, acc}$, the  $\gamma_{\rm 0, rad}$ can be calculated according to the relativistic superposition \citep{2010ASTRA...6....1W},
\begin{equation}\label{radtoacc}
\gamma_{\rm 0, rad} = [{1-(\frac{\sqrt{\Gamma^2-1}\Gamma+\sqrt{\gamma_{\rm 0, acc}^2-1}\gamma_{\rm 0, acc}}{\Gamma^2+\gamma_{\rm 0, acc}^2-1})^2}]^{-1/2}=2.3\times10^3.
\end{equation}
Finally, there are seven free parameters in our SED fitting. The minimum ${\chi}^2$ technique is used to perform the SED fits. For the radio and optical data with no reported uncertainties, we take 1$\%$ of the observed flux as the error. For the UV and X-ray data with no reported uncertainties, we take 2$\%$ of the observed flux as the error \citep{2012ApJ...752..157Z, 2014A&A...567A.135A}. The fitting result of Mrk 421 is shown in Fig.~\ref{sed}. The red dotted line represents synchrotron emission in acceleration zone. The green dashed line represents synchrotron and SSC emission in radiation zone. The yellow solid line represents the total spectrum by summation both of the acceleration zone and radiation zone emission. The model parameters used for the fitting and the values of ${\chi}^2$ are given in Table.~\ref{para}. The columns in Table.~\ref{para} are as follows: 1. source name; 2. redshift; 3. the constant injection rate in unit of $\rm cm^{-2}s^{-1}$; 4. the magnetic field in the unit of G; 5. the radius of the cross-section in unit of cm; 6. the length of the radiation zone in unit of cm; 7. time of observation in unit of s; 8. the Doppler factor; 9. $t_{\rm acc}/t_{\rm esc}$; 10. the value of ${\chi}^2$. 

From Fig.~\ref{sed}, we can see that the full-waveband observation data of Mrk 421 is reproduced well, including the flat radio spectrum. Fig.~\ref{sed} show that the emission beyond $10^{11}$Hz is dominated by the radiation zone, while the only contribution from the acceleration zone is between $10^{9}$ Hz and  $10^{11}$ Hz. This result suggests that the synchrotron radio emission ($\nu>10^9~\rm Hz$) in the acceleration zone is optically thin and the low frequency flat radio spectrum can be explained by the superposition of the two-zone synchrotron emission self-consistently. 

\begin{table*}
\begin{minipage}{185mm}
\centering
\caption{Model parameters.}
\begin{tabular}{@{}lrrrrrrrrrrrl@{}}
\hline\hline
Source name & z & $Q_0$ & $B$  & $ R $  & $L_{\rm rad}$   & $t_{\rm obs}$ & $\delta$ & $t_{\rm acc}/t_{\rm esc}$ & $\chi^2$ \\
  &  & $\rm {(cm^{-2} s^{-1})}$ & (G) &  (cm) & (cm)   &  (s) & & & \\
\hline
Mrk 421 & 0.03 & $8.90\times10^{8}$ & 0.012 & 1.41$\times$$10^{17}$ &  $1.06\times10^{18}$ &  $2.4\times10^{7}$ & 24 & 1.4 & $1.28$ \\
\hline
\end{tabular}\label{para}
\end{minipage}
\end{table*}

\subsection{The evolution of the electron energy distribution in the radiation zone}
\begin{figure}
\includegraphics[width=84mm]{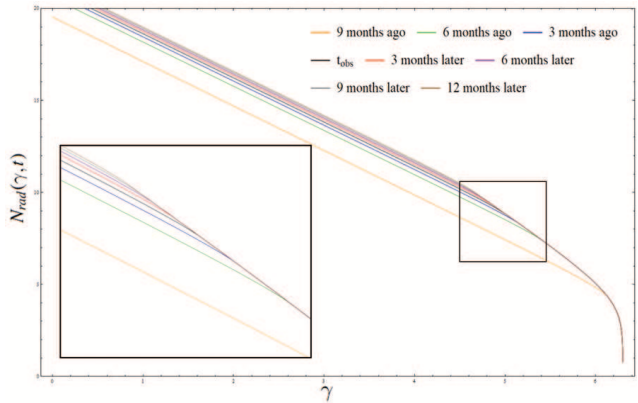}
\caption{The electron energy distribution in the radiation zone at different times. It can be seen that the electron energy distribution does not evolve significantly at large times and it is reasonable to treat our electron energy distribution that observed at $t_{\rm obs}$ as a steady state.} \label{evo}
\end{figure}

The multi-frequency data is collected during 4.5 month campaign, and shows low activity at all frequency during this period \citep{2011ApJ...736..131A}. It means that the electron energy distribution is closer to the steady state in this period. Moreover, in Fig.~\ref{sed} we can find that almost all the emission is dominated by the radiation zone. The electron energy distribution in the radiation zone can be solved by Eq.~\ref{rad} and the solution is a time-dependent function. This time-dependent two-zone model applied in this work is also used to study the multi-wavelength variabilities of some blazars \citep{2010A&A...515A..18W}. As we know, Mrk 421 is a highly variable blazar, and this work focuses on studying the origin of radio spectrum in a quiescent state. Therefore it is valuable to study the evolution of the electron energy distribution over time and figure out whether the electron energy distribution at $t_{\rm obs}$ is in steady state or in variable state. According to the parameters that constrained by the ${\chi}^2$ fitting, the model predicted SED of Mrk 421 is observed at $t_{\rm obs} = 2.4\times10^7 $seconds. The electron energy distribution in $t_{\rm obs}$, $t_{\rm obs}\pm3$ months, $t_{\rm obs}\pm$6 months, $t_{\rm obs}\pm$9 months and $t_{\rm obs}$+12 months are presented in Fig.~\ref{evo}. As can be seen, with time increases, the electron energy distribution is getting closer to the steady state. We find that the electron energy distribution has already began to be closed to the steady state at $t_{\rm obs}$-6 months in our model. This result suggest that the electron energy distribution does not evolve significantly at large times and it is reasonable to treat our electron energy distribution that observed from $t_{\rm obs}$-6 months to $t_{\rm obs}$+12 months as a steady state. This period is much longer than 4.5 month. It means that our model is consistent with the observation result during the multi-frequency campaign.

\section{Discussion and conclusion}\label{dis}

As a common property, most blazars show a flat spectrum in the low frequency radio band. For spectral index, many studies adopt $\alpha = 0$ for the low frequency radio band \citep{2000ApJ...537...80C, 2001A&A...375..739D, 2010ApJ...716...30A, 2013Ap&SS.345..345X}. A statistical study of the radio spectrum of extragalactic radio sources is presented in \citet{2011A&A...536A..15P}. Their results suggest that the distribution of the low frequency spectral indices is narrow and nearly 91$\%$ of indices are in the range $\alpha$ = -0.5 $\sim$ 0.5. As expected, their results suggest that the low frequency radio spectrum is flat with an average value of 0.06. 

As one of the challenges, one-zone leptonic model is difficult to explain the flat radio spectrum. There is, however, disagreement concerning the explanation of the flat radio spectrum. (1) It comes from a large-scale jet. However, the extended radio emission from the large-scale jet should shows a steep spectrum, but not a flat spectrum \citep{1995AJ....109.1555P, 2011ApJ...740...98M, 2015AJ....149...46Z}. (2) Some researchers believe that the flat radio spectrum is result from the superposition of many self-absorbed synchrotron components with different turnover frequencies in the inner part of jets \citep{1969ApJ...155L..71K, 1983ApJ...264..451U, 2011A&A...536A..15P}. (3) Some studies suggest that inhomogeneous models can explain the flat radio spectrum \citep{1980ApJ...235..386M, 1981ApJ...243..700K, 1985A&A...146..204G, 2012MNRAS.423..756P}.

From a phenomenal perspective, the direct contribution from one additional region or the superposition of this additional region and a compact blob is much simpler to form the flat radio spectrum. In the radio observation of AGN, the extended radio emission could have a significant contribution that may dominate over the radio emission from inner jet. However, \cite{2016Ap&SS.361..237P} finds that the radio core dominance of Mrk 421 is 0.84. It suggests that the flux of radio core emission is about one order of magnitude higher than that of the extended radio emission. Therefore, the contribution of the extended region to the observed radio spectrum is minor, and we have not considered it in this work. On the basis of the one-zone model, we further consider the contribution of the acceleration zone around the shock front to the observed SED. In this work, we present a two-zone model to fit the broadband SED of Mrk 421 in the steady state. From our fitting result, we suggest that the low frequency flat radio spectrum of Mrk 421 can be explained as a superposition of the synchrotron emission from acceleration zone and radiation zone. Among them, the contribution from the acceleration zone comes from its steep radio spectrum, and the contribution from the radiation zone comes from its radio spectrum with $\alpha_{\rm rad} = -1/3$. As the spectral index of the steep radio spectrum in the acceleration zone is the same as the steep radio spectrum in the radiation zone, we can study it through the observed radio spectrum. \cite{2012A&A...541A.160G} study the low and high frequency radio spectrum with power law to estimate their spectral indices of 105 blazars. The average value of the high frequency spectral indices ($\alpha_{\rm HF}$, for $\nu > \nu_{\rm break}$) of FSRQs and BL Lacs are $\alpha_{\rm HF} = 0.73\pm0.04$ and $\alpha_{\rm HF} = 0.51\pm0.07$, respectively.  The $\alpha_{\rm HF} = 0.73\pm0.04$ for FSRQs emerges naturally from the simplest first-order Fermi acceleration mechanism and the $\alpha_{\rm HF} = 0.51\pm0.07$ for BL Lacs is in agreement with the simplest diffusive shock acceleration models \citep{1998PhRvL..80.3911B, 2007Ap&SS.309..119R}. By superimposing the spectrum from the radiation zone with $\alpha_{\rm rad}$ = -1/3 and the steep spectrum from the acceleration zone that can be explained by the acceleration mechanisms, we can obtain the low frequency flat radio spectrum with spectral indices($\alpha_{\rm LF}$, for $\nu < \nu_{\rm break}$) of $\alpha_{\rm LF} \approx 0.2$ for FSRQs and $\alpha_{\rm LF} \approx 0.08$ for BL Lacs. This result is consistent with the statistical average value of 0.1 that presented by \cite{2012A&A...541A.160G} which suggests that our model can explain the low frequency flat radio spectrum self-consistently.

As introduced in Sect.~\ref{Modeling}, when considering the upstream electrons moving in the opposite direction, $\gamma_{\rm 0,rad}$ is much higher than $\gamma_{\rm 0,acc}$. Moreover, we suggest that the difference between $\gamma_{\rm 0, rad}$ and $\gamma_{\rm 0, acc}$ may implies some physical mechanisms worked in the shock. The energy dissipation mechanisms operating at the shock fronts do introduce a particular characteristic (injection) energy scale, below which the particles are not freely able to cross the shock front and enter the radiation zone \citep{2011ApJ...727..129A}. Such a scale depends critically on the ratio of number of leptons to protons in the shocked plasma ($q$) and the fraction of which the shock energy is transferred to the acceleration of leptons ($\epsilon_{\rm e}$)\citep{2016ApJ...828...13I}:
\begin{equation}
E_{\rm c} = \Gamma m_{\rm p} c^2 \frac{\epsilon_{\rm e}}{q} \frac{p-2}{p-1},
\end{equation}
where $E_{\rm c} = \gamma_{\rm 0, rad}m_{\rm e}c^2$, $p\approx2.58$ is the spectral index derived in the modeling, and $\Gamma = \delta = 24$\footnote{For the relativistic jet close to the line of sight in blazars with a viewing angle of $\theta \lesssim 1/\Gamma$, we have $\delta=[\Gamma(1-\beta \rm cos\theta)]^{-1}\approx \Gamma$.}. We consider that jet is electrically neutral, and the number for electrons and protons is approximately equal ($q$ $\approx$ 1), then we can get the value $\epsilon_{\rm e} \approx$ 0.14. It means that 14$\%$ of shock energy goes into electron acceleration which is consistent with the value (10$\%$) that obtained in \cite{2011ApJ...727..129A} and \cite{2016ApJ...828...13I}.

As a conclusion, we propose an alternative interpretation of the low frequency flat radio spectrum of Mrk 421. The fitting result shows that almost all the emission is still dominated by the radiation zone. However, in the radio band, both the acceleration zone and the radiation zone can contribute to the radio flux. Our model also suggest that the flat radio emission of the jet in Mrk 421 is originated from a compact cylindrical region rather than a large scale region. On the other hand, the two-zone model that presented in this paper still requires improvements. For example, the length of the radiation zone is sufficient long to consider the energy conservation, bulk acceleration and electron continuity along the jet. Among them, only the electron continuity from acceleration zone to the tail of radiation zone is taken into account in the two-zone model. In addition, two-zone model can only reproduce a relatively narrow region of a flat radio spectrum. In order to explain the low frequency flat radio spectrum over a larger and varying range generally, a more complex radiation model should be expected.



\section*{Acknowledgements}
We thank the anonymous referee for insightful comments and constructive suggestions. This work is supported by the Ph.D. Research start-up fund of Zhejiang Normal University (Grant No. YS304320082).

\label{lastpage}

\clearpage

\end{document}